\documentclass[letter,prl,twocolumn,superscriptaddress,showpacs]{revtex4-1}

\usepackage{amsmath}
	\usepackage{makeidx}
	\usepackage{amsfonts}
	\usepackage[ansinew]{inputenc}
	\usepackage[usenames,dvipsnames]{pstricks}
	\usepackage{subfigure}
	\usepackage{epsfig}
	\usepackage{pst-grad} 
	\usepackage{pst-plot} 
	\usepackage[colorlinks,hyperindex]{hyperref}
	\usepackage{mathrsfs}
	\usepackage{graphicx}
  \usepackage{bm}
  \usepackage{amssymb}
	\hypersetup
	{
		colorlinks,%
		citecolor=green,%
		linkcolor=blue,%
		urlcolor=red,%
	}

	\newcommand{\subfigimg}[3][,]{%
  \setbox1=\hbox{\includegraphics[#1]{#3}}
  \leavevmode\rlap{\usebox1}
  \rlap{\hspace*{10pt}\raisebox{\dimexpr\ht1-2\baselineskip}{#2}}
  \phantom{\usebox1}
	}
  \newcommand{\squeezeup}{\vspace{-2.5mm}}
\makeindex
\begin{document}

\title{Nonlinear Alfv\'en waves in extended magnetohydrodynamics }

\author{Hamdi M. Abdelhamid}
\email{hamdi@ppl.k.u-tokyo.ac.jp}
\affiliation{Graduate School of Frontier Sciences, University of Tokyo, Kashiwanoha, Kashiwa, Chiba 277-8561, Japan}
\affiliation{Physics Department, Faculty of Science, Mansoura University, Mansoura 35516, Egypt}
\author{Zensho Yoshida}
\email{yoshida@k.u-tokyo.ac.jp}
\affiliation{Graduate School of Frontier Sciences, University of Tokyo, Kashiwanoha, Kashiwa, Chiba 277-8561, Japan}
\date{\today }

\begin{abstract}
Large-amplitude Alfv\'en waves are observed in various systems in space and laboratories,
demonstrating an interesting property that the wave shapes are stable even in the nonlinear regime.
The ideal magnetohydrodynamics (MHD) model predicts that an Alfv\'en wave
keeps an arbitrary shape constant when it propagates on a homogeneous ambient magnetic field.
However, such arbitrariness is an artifact of the idealized model that omits the dispersive effects.
Only special wave forms, consisting of two component sinusoidal functions, can maintain the shape;
we derive fully nonlinear Alfv\'en waves by an extended MHD model that includes both the Hall and electron inertia effects.
Interestingly, these ``small-scale effects'' change the picture completely;
the large-scale component of the wave cannot be independent of the small scale component,
and the coexistence of them forbids the large scale component to have a free wave form.
This is a manifestation of the nonlinearity-dispersion interplay, which is somewhat different from that of solitons. 
\end{abstract}
\maketitle
Alfv\'en waves are the most typical electromagnetic phenomena in magnetized plasmas. In particular, nonlinear Alfv\'en waves deeply influence various plasma regimes in laboratory as well as in space, which have a crucial role in plasma heating \cite{Hasegawa74, Nature11}, turbulence\cite{Sridhar94,Sridhar95,Boldyrev06}, reconnection\cite{Rogers01}, etc. 
As an interesting property of the Alfv\'en waves, the amplitudes as well as wave forms
are totally arbitrary when they propagate on a homogenous ambient magnetic field \cite{Alfven, Alfven1}.
In fact, we often observe large-amplitude Alfv\'en waves in orderly propagation (for example\,\cite{Alfven2}).
To put it in theoretical language, the set $A$ of Alfv\'en waves after an appropriate transformation (see \cite{mahajan98}), is a closed linear subspace,
i.e., every linear combination of the members of $A$ gives solution to the fully nonlinear wave equation.
Needless to say, the set of general solutions to a linear equation is, by definition, a linear subspace. 
However, it is remarkable that the nonlinear MHD equation
has such a linear subspace $A$ of solutions.

Here we investigate the underlying mechanism producing such solutions in the context of a more accurate framework, generalized MHD.
When we take into account dispersion effects (we consider both ion and electron inertial effects \cite{Morrison,hamdi15}),
the wave forms are no longer arbitrary (remember that the ideal MHD model is dispersion free).  
Yet, we find that the generalized MHD system has a linear subspace of nonlinear solutions.
The Casimir invariants of the system is the root cause of this interesting property\cite{Yoshida12}.

We start by reviewing how the nonlinear Alfv\'en waves are created in ideal MHD;
we put the problem in the perspective of Hamiltonian mechanics.
We then formulate the generalized MHD system in a Hamiltonian form.
Via constructing equilibrium solutions (so-called Beltrami equilibriums) on Casimir leaves, 
we derive nonlinear wave solutions.
The dispersion relation is exactly that of the linear theory, while the wave amplitude may be arbitrarily large.
The wave function is composed of two components bearing distinct length scales. 

Here we consider an ideal MHD plasma obeying
\begin{eqnarray}
\frac{\partial \rho }{\partial t}&=& -\nabla\cdot(\bm{V}\rho ),
\label{ideal_MHD-0}\\
\frac{\partial\bm{V}}{\partial t} &=& -(\nabla\times\bm{V})\times\bm{V}+\rho^{-1}(\nabla\times\bm{B})\times\bm{B} 
\nonumber \\
& & ~~~~~~~~~~~~~~~~~ -\nabla(V^2/2 + h),
\label{ideal_MHD-1}
\\
\frac{\partial \bm{B}}{\partial t} &=& \nabla\times(\bm{V}\times\bm{B}) ,
\label{ideal_MHD-2}
\end{eqnarray}
where $\rho$ is the density, $\bm{V}$ is the velocity, $\bm{B}$ is the magnetic field, and $h\left(\rho\right)$ is the total enthalpy.
All variables are written in the standard Alfv\'en units.
The ideal MHD equation has a noncanonical Hamiltonian structure\,\cite{Morrison-Green}.

Hamiltonian and Poisson operator of MHD
\begin{equation}
H =\int_{\Omega}\left\{\rho\left(\frac{\left|\bm{V}\right|^{2}}{2}+U\left(\rho\right)\right)+\frac{\left|\bm{B}\right|^{2}}{2}
\right\} d^{3}x,
\label{MHD-Hamiltonian}
\end{equation}
\small
\begin{eqnarray}
\label{44}
\mathcal{J}_{MHD} = \left(
\begin{array}{ccc}
	0&-\nabla\cdot&0\\
	\\
	-\nabla&-\rho^{-1}\left(\nabla\times\bm{V}\right)\times\circ&\rho^{-1}\left(\nabla\times\circ\right)\times\bm{B}\\
	\\
0&\nabla\times\left(\circ\times\rho^{-1}\bm{B}\right)&0
\end{array}
\right).\nonumber\\
\end{eqnarray}
\normalsize
The existence of Casimir invariants is the signature of the noncanonicality,
by which the orbits in the phase space are restricted to stay on the Casimir leaves (the level-sets of the Casimir invariants).
The equilibrium points are, then, the stationary points of the Hamiltonian (energy) on the Casimir leaves.
The cross helicity $C_{\mathrm{cross}}=\int_\Omega \bm{V}\cdot\bm{B}\,d^3x$ ($\Omega$ is the total volume of the plasma) 
is one of the Casimir invariants of MHD,
which is relevant to the present purpose of constructing nonlinear Alfv\'en waves.

Minimizing the Hamiltonian $H$ with the constraint on $C_{\mathrm{cross}}$, we obtain 
\begin{equation}
\bm{V}=\pm\bm{B}.
\label{ideal_MHD_equilibrium}
\end{equation}
Evidently, every $\bm{B}$,
being combined with $\bm{V}=\pm\bm{B}$,
is an equilibrium ($\partial_t=0$) solution of the ideal incompressible MHD equations.

We can convert these equilibrium stats to Alfv\'en waves propagating on a homogeneous ambient magnetic field $\bm{B}_0$
(which can be arbitrarily chosen)\,\cite{Yoshida12}.
Let us rewrite $\bm{B}$ and $\bm{V}=\pm\bm{B}$ as
\begin{equation}
\bm{B} = \bm{B}_0 + \bm{b}, \quad \bm{V}= \pm\bm{B}_0 + \bm{v}.
\label{ideal_MHD_Alfven}
\end{equation}
Boosting the coordinate $\bm{x} \rightarrow \bm{x} \mp \bm{B}_0 t$,
we find that the decomposed component (which is the wave component) satisfies
\begin{eqnarray}
\frac{\partial\bm{v}}{\partial t} &=& -(\nabla\times\bm{v})\times\bm{v}+(\nabla\times\bm{b})\times(\bm{b}+\bm{B}_0)
\nonumber \\
& &~~~~~~~~~~~~~~~~~~~ -\nabla(V^2/2 + P),
\label{ideal_MHD-1'}
\\
\frac{\partial \bm{b}}{\partial t} &=& \nabla\times[\bm{v}\times(\bm{b}+\bm{B}_0)] ,
\label{ideal_MHD-2'}
\end{eqnarray}
which are exactly the Alfv\'en wave equations with an ambient field $\bm{B}_0$.
Notice that the wave component $\bm{b}$ and $\bm{v}$ propagate with the Alfv\'en velocity $\pm\bm{B}_0$.

Now we will proceed to extend our analysis to a generalized MHD. We start with the dimensionless extended MHD equation \cite{Morrison,hamdi15}, which is composed of 
\begin{eqnarray}
\label{1}
\frac{\partial \rho}{\partial t}&=&-\nabla\cdot\left(\rho\bm{V}\right),
\\
\label{2}
\frac{\partial \bm{V}}{\partial t}&=&-\left(\nabla\times\bm{V}\right)\times\bm{V}+\rho^{-1} \left(\nabla\times\bm{B}\right)\times\bm{B}^{\ast}\nonumber \\
& &~~~~-\nabla\left(h+V^{2}/2+d^{2}_{e}\left(\nabla\times\bm{B}\right)^{2}/2\rho^{2}\right),
\\
\label{3}
\frac{\partial \bm{B}^{\ast}}{\partial t}&=&\nabla\times\left(\bm{V}\times\textbf{B}^{\ast}\right)- \nabla\times\left(\rho^{-1} \left(\nabla\times\bm{B}\right)\times\bm{B}^{\ast}\right)\nonumber \\
& &~~~~+d^{2}_{e} \nabla\times\left(\rho^{-1} \left(\nabla\times\bm{B}\right)\times\left(\nabla\times\bm{V}\right)\right),
\end{eqnarray}
where
\begin{eqnarray}
\label{4}
\bm{B}^{\ast}&=&\bm{B}+d^{2}_{e}\nabla\times\rho^{-1}\left(\nabla\times\bm{B}\right),
\end{eqnarray}
$d_{e}=c/(\omega_{pe}d_{i})$ is the normalized electron skin depth normalized to ion skin depth $d_{i}=c/\left(\omega_{pi}\right)$, $\omega_{pe}$ and $\omega_{pi}$ are the electron and ion plasma frequencies and $c$ is the speed of light. Notice that for simplicity, barotropic pressure assumption is used here. The above system of equations are normalized in Alfv{\'e}nic units defined as follow: the magnetic field is normalized to the ambient filed $B_{0}$, the velocity to the Alfv{\'e}n speed ($V_{A}=B_{0}/\sqrt{\mu_{\circ}\rho}$ ($\mu_{\circ}$the vacuum permeability)), time to the ion gyroperiod $\omega^{-1}_{ci}$, and the space variables to the ion skin depth $d_{i}$.

Equations \eqref{1},\eqref{2} and \eqref{3} with the total energy 
\begin{eqnarray}
\label{6}
\mathscr{H}=\int_{\Omega}\left\{\rho\left(\frac{\left|\bm{V}\right|^{2}}{2}+U\left(\rho\right)\right)+\frac{\bm{B}\cdot\bm{B}^{\ast}}{2}
\right\} d^{3}x,
\end{eqnarray}
and Poisson operator 
\small
\begin{eqnarray}
\label{Poisson}
\mathcal{J}_{XMHD} = \left(
\begin{array}{ccc}
	0&-\nabla\cdot&0\\
	\\
	-\nabla&-\frac{\left(\nabla\times\textbf{V}\right)\times\circ}{\rho}&\frac{\left(\nabla\times\circ\right)\times\textbf{B}^{\ast}}{\rho}\\
	\\
0&\nabla\times\frac{\left(\circ\times\textbf{B}^{\ast}\right)}{\rho}&\bigg[-\nabla\times\left(\frac{\left(\nabla\times\circ\right)\times\textbf{B}^{\ast}}{\rho}\right)\\&&+d^{2}_{e}\nabla\times\left(\frac{\left(\nabla\times\circ\right)\times\left(\nabla\times\textbf{V}\right)}{\rho}\right)\bigg]
\end{array}
\right),
\nonumber\\
\end{eqnarray}
\normalsize
constitute a noncanonical Hamiltonian system in which the phase space is spanned by the dynamical variables
$\rho, \bm{V}$, and $\bm{B}^{\ast}$. Notice that, the details of derivation of the Poisson bracket and the proof of the Jacobi identity were given in \cite{hamdi15}. 

The extended MHD has three independent Casimirs:
\begin{eqnarray}
\label{c2}
C_{1}&=&\int_{\Omega}\bm{B}^{\ast}\cdot\left(\bm{V}-\frac{1}{2d^{2}_{e}}\bm{A}^{\ast}\right) d^{3}x,
\\
\label{c3}
C_{2}&=&\frac{1}{2}\int_{\Omega}\big[\bm{B}^{\ast}\cdot\bm{A}^{\ast}+d^{2}_{e}\bm{V}\cdot\left(\nabla\times\bm{V}\right)\big] d^{3}x,
\\
\label{c1}
C_{3}&=&\int_{\Omega}\rho~d^{3}x.
\end{eqnarray}
To construct the Beltrami equilibria, we start from the energy-Casimir functional of the extended MHD system, which reads as
\begin{eqnarray}
\label{ec}
\mathscr{H}_{\mu}\left(u\right)=\mathscr{H}\left(u\right)-\sum^{3}_{n=1}\mu_{n} C_{n}\left(u\right).
\end{eqnarray}
The critical points on the Casimir leaves are found by setting $\partial_{u}\mathscr{H}_{\mu}=0$ which yields
\begin{eqnarray}
\label{ec1}
\nabla\times\bm{B}&=&\mu_{1}\nabla\times\bm{V}+\left(\mu_{2}-\frac{\mu_{1}}{d^{2}_{e}}\right)\bm{B}^{\ast},
\end{eqnarray}
\begin{eqnarray}
\label{ec2}
\rho\bm{V}&=&\mu_{1}\bm{B}^{\ast}+\mu_{2}d^{2}_{e}\nabla\times\bm{V},
\end{eqnarray}
\begin{eqnarray}
\label{ec3}
\frac{V^{2}}{2}&+&h\left(\rho\right)+d^{2}_{e}\frac{\left(\nabla\times\bm{B}\right)^{2}}{2\rho^{2}}-\mu_{3}=0,
\end{eqnarray}
where $\mu_{1}$, $\mu_{2}$ and $\mu_{3}$ are Lagrange multipliers. Notice that \eqref{ec3} is Bernoulli's equation. Now, consider the incompressible flow $\left(\nabla\cdot\bm{V}=0\right)$ with a constant mass density $\rho=1$. 
Then, combining \eqref{ec1} and \eqref{ec2} with the aid of \eqref{4}, we get the triple curl Beltrami equation, 
\begin{eqnarray}
\label{3curl}
\nabla\times\nabla\times\nabla\times\bm{B}-\eta_{1}\nabla\times\nabla\times\bm{B}+\eta_{2}\nabla\times\bm{B}-\eta_{3}\bm{B}=0,\nonumber\\
\end{eqnarray}
 where \begin{eqnarray*}
\eta_{1}&=&\left(2-\frac{\mu_{1}}{d^{2}_{e}\mu_{2}}\right)/\Delta,
\\
\eta_{2}&=&\left(\mu_{2}+\frac{1}{d^{2}_{e}\mu_{2}}-\frac{\mu_{1}}{d^{2}_{e}}\left(1+\frac{\mu_{1}}{\mu_{2}}\right)\right)/\Delta,
\\
\eta_{3}&=&\left(1-\frac{\mu_{1}}{d^{2}_{e}\mu_{2}}\right)/\left(d^{2}_{e}\Delta\right),
\end{eqnarray*}
\begin{eqnarray*}
\Delta=d^{2}_{e}\left[\mu_{2}-\frac{\mu_{1}}{d^{2}_{e}}\left(1+\frac{\mu_{1}}{\mu_{2}}\right)\right].
\end{eqnarray*}
The general solution of \eqref{3curl} can be expressed in terms of a single Beltrami fields $\bm{G}_{l} \left(l=0,1,2\right),$ such that 
\begin{align*}
\begin{split}
\left(curl-\lambda_{l}\right)\bm{G}_{l}&=0~~~~~~\left(in~~\Omega\right),\\
\textbf{n}\cdot\bm{G}_{l}&=0~~~~~~\left(on~~\Omega\right)
\end{split}
\end{align*}
for more details see \cite{mahajan98, Yoshida02}. Then, \eqref{3curl} can be written as
\begin{eqnarray}
\label{3curl2}
\left(curl-\lambda_{0}\right)\left(curl-\lambda_{1}\right)\left(curl-\lambda_{2}\right)\bm{B}=0,
\end{eqnarray}
where the eigenvalues $\lambda_{0}$, $\lambda_{1}$ and $\lambda_{2}$ are given by
\begin{align}
\begin{split}
\label{eigen}
\lambda_{0}+\lambda_{1}+\lambda_{2}&=\eta_{1},\\
\lambda_{0}\lambda_{1}+\lambda_{1}\lambda_{2}+\lambda_{2}\lambda_{0}&=\eta_{2},\\
\lambda_{0}\lambda_{1}\lambda_{2}&=\eta_{3}.
\end{split}
\end{align}
Now constructing the general solution, which is the linear combination of three eigenfunctions given as,
\begin{eqnarray}
\label{gsolb}
\bm{B}=a_{0}\bm{G}_{0}+a_{1}\bm{G}_{1}+a_{2}\bm{G}_{2},
\end{eqnarray}
where $a_{l}$'s are arbitrary constants. Substituting in \eqref{ec1} and \eqref{ec2}, the corresponding flow is given by
\begin{multline}
\label{gsolv}
\bm{V}=\left[\sigma\left(1+d^{2}_{e}\lambda^{2}_{0}\right)+\frac{d^{2}_{e}\mu_{2}}{\mu_{1}}\lambda_{0}\right]a_{0}\bm{G}_{0}\\+\left[\sigma\left(1+d^{2}_{e}\lambda^{2}_{1}\right)+\frac{d^{2}_{e}\mu_{2}}{\mu_{1}}\lambda_{1}\right]a_{1}\bm{G}_{1}\\+\left[\sigma\left(1+d^{2}_{e}\lambda^{2}_{2}\right)+\frac{d^{2}_{e}\mu_{2}}{\mu_{1}}\lambda_{2}\right]a_{2}\bm{G}_{2},
\end{multline}
where $\sigma=\mu_{1}+\mu_{2}\left(1-\frac{d^{2}_{e}\mu_{2}}{\mu_{1}}\right)$.

Now, setting one of the Beltrami eigenvalues equal to zero $\left(\lambda_{0}=0\right)$(which implies that the corresponding eigenfunction is a harmonic field), yields a special class of Beltrami solutions, see\cite{Yoshida12}. Let $\left(\lambda_{0}=0\right)$, two consequences immediately follow from \eqref{eigen},
\begin{align}
\mu_{1}=d^{2}_{e}\mu_{2}&=\mu,\\
\begin{split}
\label{eigen2}
\lambda_{1}+\lambda_{2}&=\eta_{1},\\
\lambda_{1}\lambda_{2}&=\eta_{2}.
\end{split}
\end{align}
Now, solving \eqref{eigen2} yields
\begin{eqnarray}
\label{lambda}
\lambda_{\pm}=\frac{1}{2\mu d^{2}_{e}}\left[-1\pm\sqrt{1-4d^{2}_{e}\left(\mu^{2}-1\right)}\right],
\end{eqnarray}
where we chose $\lambda_{+}=\lambda_{1}$ and  $\lambda_{-}=\lambda_{2}$. Under this conditions the general flow solution becomes
\begin{eqnarray}
\label{gsolv2}
\bm{V}=\mu a_{0}\bm{G}_{0}+ \frac{1}{\mu}\left(a_{1}\bm{G}_{1}+a_{2}\bm{G}_{2}\right).
\end{eqnarray}
Based on the geometry ($xyz-plane/space$), the eigenfunctions $\bm{G}_{1}$ and $\bm{G}_{2}$ are naturally a sinusoidal functions. To satisfy the single Beltrami condition, the eigenfunctions are given in the form of a circularly polarized wave;  
\begin{eqnarray*}
\bm{G}_{1} = \left(
\begin{array}{c}
	\sin\left(\lambda_{1}z\right)
	\\
	\cos\left(\lambda_{1}z\right)
	\\
0
\end{array}
\right),~~~~~~~
\bm{G}_{2} = \left(
\begin{array}{c}
	\sin\left(\lambda_{2}z\right)
	\\
	\cos\left(\lambda_{2}z\right)
	\\
0
\end{array}
\right).\nonumber
\end{eqnarray*}
An immediate generalization to a more complex Beltrami field, so called $ABC$ flow is possible [cf.\cite{mahajan2005}].
These $\bm{V}$ and $\bm{B}$ have oscillatory amplitudes, thus the Bernoulli condition \eqref{ec3} demands a non-constant $h\left(\rho\right)$. We assume a large enough sound velocity to keep $\rho$ be almost constant as we have assumed in \eqref{3curl} . Considering density perturbation will be found in a forthcoming article; the methodology can be found in \cite{Yoshida12}. On the other hand, Beltrami solutions \eqref{gsolb} and \eqref{gsolv2} imply that the magnetic field and the flow velocity are not necessarily aligned, unless $\mu=\pm1$. 
\begin{figure}
\centering
\begin{tabular}{@{}p{0.85\linewidth}@{}}
\subfigimg[width=\linewidth]{(a)}{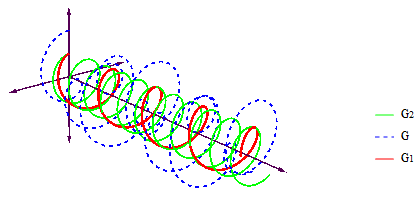}\\
\subfigimg[width=\linewidth]{(b)}{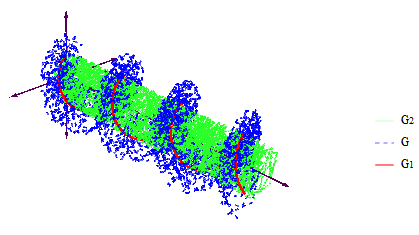}
\end{tabular}
\caption{(Color online) The profiles of the eigenfunctions $\bm{G}_{1}$, $\bm{G}_{2}$ and their superposition $\bm{G}=a_{1}\bm{G}_{1}+a_{2}\bm{G}_{2}$ for $\mu=2$, $a_{1}=a_{2}=1$; (a)$d_{e}=0.26$ and (b)$d_{e}=10^{-8}$.}
\label{Fig:wave1}
\squeezeup
\vspace{-2mm}
\end{figure}
Additionally, we observe that the solutions are expressed as a combination of three Beltrami eigenfunctions $\bm{G}_{l}$, in which two of them have a large scale (compared with the electron skin depth $d_{e}$), whereas the third is in scale hierarchy of $d_{e}$. Since in the Hall MHD limit $d_{e}\rightarrow0$, one of the eigenvalues ($\lambda_{+}\rightarrow \left(1-\mu^{2}\right)/\mu$) is finite, whilst the other ($\lambda_{-}\rightarrow-\infty$) is singular and therefore the corresponding eigenstate $\bm{G}_{2}$ is divergent; see Fig.\ref{Fig:wave1}. This singularity can be removed by setting the arbitrary constant ($a_{2}$) associated with the divergent eigenstate ($\bm{G}_{2}$) to zero. The physical insight and the derivation of the condition that remove the singular part of the solutions will be the subject of further publication.

To examine the propagation of the wave component, we assume that $\bm{G}_{0}$ serve as an ambient field. Now, setting $\bm{G}_{0}=\widehat{\bm{e}}_{z}$ and $ a_{0}=1$, ( $\bm{G}_{0}$  represents the normalized ambient magnetic field). From \eqref{gsolv2} the corresponding ambient flow is $\bm{V}_{0}=\mu~\widehat{\bm{e}}_{z}$. 
The magnetic and flow fields become
\begin{eqnarray}
\label{ambi1}
\bm{B}&=&\bm{b}+\widehat{\bm{e}}_{z},~~~~~~~~~\bm{V}=\bm{v}+\mu \widehat{\bm{e}}_{z},
\\
\text{where}~~~~~~~~
\label{lincs11}
\bm{b}&=&\mu\bm{v}.
\end{eqnarray}
Let us show explicitly that the Beltrami solution \eqref{ambi1}-\eqref{lincs11} can be modified to wave solution by boosting the coordinate. The Beltrami solution is the stationary solution satisfying
\begin{eqnarray}
\label{inc1}
0=\nabla\times\big[\left(\bm{V}-\nabla\times\bm{B}\right)\times\bm{B}^{\ast}\big],
\\
\label{inc2}
0=\nabla\times\big[\bm{V}\times\left(\bm{B}^{\ast}+\nabla\times\bm{V}\right)\big],
\\
\label{inc3}
\nabla\cdot\bm{V}=0, 
\\
\label{inc4}
\nabla\cdot\bm{B}=0,~~~~~~~~~\nabla\cdot\bm{B}^{\ast}=0.
\end{eqnarray}
Transforming the system under Galilean-boost yields the new coordinates:
\begin{eqnarray*}
\label{trans}
\left(x,y,z\right)\longmapsto\left(x,y,\xi\right):=\left(x,y,z-\mu t\right).
\end{eqnarray*}
where $t\mapsto\tau:=t$ and $z\mapsto\xi:=z-\mu t$. The transformations of derivatives with respect to the coordinates are
\begin{eqnarray*}
\nabla_{x,y,z}\mapsto\widetilde{\nabla}_{x,y,\xi},~~~~~~~~~\frac{\partial}{\partial t}\mapsto\frac{\partial}{\partial\tau}-\mu\frac{\partial}{\partial\xi}
\end{eqnarray*}
where for 3-vector $\bm{X}$ (with $\nabla\cdot\bm{X}=0$), $-\mu\frac{\partial\bm{X}}{\partial\xi}=\nabla\times\left(\mu\widehat{\bm{e}}_{z}\times\bm{X}\right)$ is true.
Using \eqref{ambi1}, equations \eqref{inc1} and \eqref{inc2} can be boosted in the new coordinates into
\begin{eqnarray}
\label{inc5}
\frac{\partial \bm{B}^{\ast}}{\partial \tau}&=&\widetilde{\nabla }\times\big[\left(\bm{v}-\widetilde{\nabla}\times\bm{B}\right)\times\bm{B}^{\ast}\big] ,
\\
\label{inc6}
\frac{\partial \left(\bm{B}^{\ast}+\widetilde{\nabla}\times\bm{v}\right)}{\partial \tau}&=&\widetilde{\nabla}\times\big[\bm{v}\times\left(\bm{B}^{\ast}+\widetilde{\nabla}\times\bm{v}\right)\big],
\end{eqnarray}
which are the Alfv\'en wave equations with  a homogeneous ambient field $\bm{B}_{0}=\widehat{\bm{e}}_{z}$. Thence, on the boosted frame, the fluctuated parts of the previous stationary solution appears as propagating waves, which forms exact solution of the incompressible extended MHD equations. Here, we notice that the wave components are a superposition of two Beltrami eigenfunctions, which implies that only a definite wave functions (sinusoidal functions), can propagate with a fixed shape.  
Further, the phase velocity here is given by $\mu$, which from \eqref{eigen2} may be written as
\begin{eqnarray}
\label{mu}
\mu_{\pm} =\frac{1}{\left(1+d^{2}_{e}k^{2}\right)}\left[-\frac{k}{2}\pm\sqrt{\frac{k^{2}}{4}+\left(1+d^{2}_{e}k^{2}\right)}\right],
\end{eqnarray}
where the eigenvalue $k:=\lambda_{1}$ or $\lambda_{2}$, serve as the wave number. 
Then, the corresponding circularly polarized wave dispersion relation $\left(\omega=-\mu\left(\widehat{\bm{e}}_{z}\cdot \bm{k}\right)\right)$, which in the case $\bm{k}=k~\widehat{\bm{e}}_{z}$ reads as   
\begin{eqnarray}
\label{dis}
\omega_{\pm} =\frac{-k}{\left(1+d^{2}_{e}k^{2}\right)}\left[-\frac{k}{2}\pm\sqrt{\frac{k^{2}}{4}+\left(1+d^{2}_{e}k^{2}\right)}\right].
\end{eqnarray}
\begin{figure}
    \centering
    \subfigure[]
    {
       \includegraphics[width=2.4in]{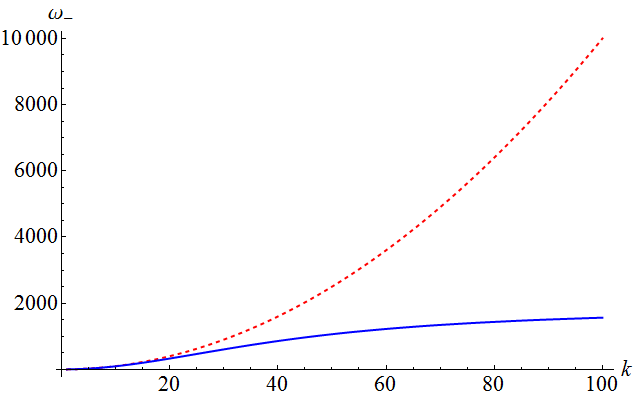}
    }
  \quad
    \subfigure[]
    {
        \includegraphics[width=2.4in]{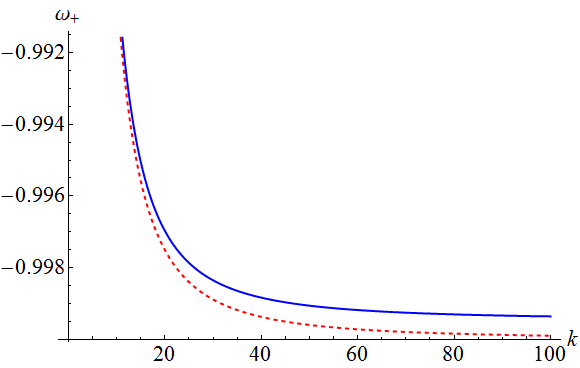}
   }
    \caption{(Color online) Normalized dispersion relation profiles for $d_{e}=0$ (dashed-red) and $d_{e}=0.0233$ (blue); (a) $\left(\omega_{-}\right)$  and (b) $\left(\omega_{+}\right)$ .}
\label{Fig1}
\squeezeup
\vspace{-2mm}
\end{figure}
which represents the dispersion relation of the fully nonlinear wave solutions. In the limit $\left(d_{e}\rightarrow 0\right)$, \eqref{dis} is reduced to the dispersion of exact solution of Hall MHD \cite{mahajan2005,Yoshida12}. We also observe that the inclusion of the electron inertia effect not only modifies waves modes, remove the singularities associated with the exact solution of Hall MHD, but also captured more of the physics of the full two-fluid model; see Figs.\ref{Fig1}. 
The previous sentence can be clarified by studying the extreme limits of equation \eqref{dis}:
\begin{enumerate}
	\item 
 for $k<<1$,
  
$
\mu_{\pm}\rightarrow \pm1,~~~~~~~\omega_{\pm}\rightarrow \mp k,
$

which represent the shear Alfv\'en wave in the ideal MHD limit.
	\item 
for $k>1$ but $d^{2}_{e} k^{2}<<1$, 
 
$
\mu_{+}\rightarrow 1/k,~~~~~~~~~\omega_{+}\rightarrow -1,\\
\mu_{-}\rightarrow -k,~~~~~~~\omega_{-}\rightarrow k^{2},
$

$\omega_{+}$ represents here ion gyrofrequency, whilst $\omega_{-}$ represents whistler wave.  
	\item 
for $k>>1$ and $d^{2}_{e} k^{2}>>1$, 

$
\mu_{\pm}\rightarrow \theta_{\pm}/k,~~~~~~~\omega_{\pm}\rightarrow -\theta_{\pm},
$

where $\theta_{\pm}= \left(-1\pm\sqrt{1+4d^{2}_{e}}\right)/2d^{2}_{e}$ are constants ($\theta_{-}$ approximates the normalized electron  gyrofrequency, $\theta_{+}$ approximates the normalized ion gyrofrequency). An important feature here is the dependence of $\theta$ on the dimensionless electron skin depth, which represent a direct relation between the electron skin depth and ion(electron) gyrofrequency.
\squeezeup
\end{enumerate}

In summary, we have given exact wave solutions of the fully nonlinear extended MHD system. 
The solutions consist of two Beltrami eigenfunctions with different length scales (macroscopic and microscopic). In contrast to the ideal MHD predictions, inclusion of the small scale effects (Hall and electron inertia effects) caused only two component sinusoidal wave functions to preserve its shape.
As shown in \eqref{lincs11}, the partition of the wave energy between $\bm{b}$ and $\bm{v}$ is determined by the phase velocity $\mu$ that is a function of $k$ (wave number). 
We can observe that, for $k>1$ only whistler wave has a magnetic energy more than the kinetic energy $\left(\bm{b}\rightarrow k\bm{v}\right)$, otherwise the kinetic energy is the dominant $\left(\bm{v}\rightarrow k\bm{b}\right)$.
This is also in a marked contrast to the ideal Alfv\'en waves in which the wave energy is equally partitioned by $\bm{b}$ and $\bm{v}$.

\begin{acknowledgments}
The authors thank Prof. P. J. Morrison, and Dr. E. Tassi for many useful discussions. H. M. Abdelhamid would like to thank the Egyptian Ministry of Higher Education for supporting his research activities. This work was partly supported by JSPS KAKENHI Grant Number 23224014 and 15K13532.
\end{acknowledgments}

\end{document}